\documentclass[aps,prb,reprint,preprintnumbers,showpacs,showkeys,superscriptaddress]{revtex4-1}
\usepackage{amsmath,amssymb,mathrsfs,graphicx, braket}
\usepackage{multirow}
\usepackage[colorlinks=true,citecolor=blue,linkcolor=blue]{hyperref}

\newcommand{ \SG }{\textrm{SG}}
\newcommand{ \oned }{\textrm{1D}}
\newcommand{ \nmin }{\nu_{\textrm{min}}}
\newcommand{\sg}[1]{{\bf\em{#1}}}

\newcommand{ \mG }{\mathcal{G}}

\usepackage[all,cmtip]{xy}

\begin{document}
\title{Filling constraints for spin-orbit coupled insulators in symmorphic and non-symmorphic crystals}

\author{Haruki Watanabe}
\affiliation{University of California, Berkeley, California 94720.}

\author{Hoi Chun Po}
\affiliation{University of California, Berkeley, California 94720.}

\author{Ashvin Vishwanath}
\affiliation{University of California, Berkeley, California 94720.}
\affiliation{Materials Science Division, Lawrence Berkeley National Laboratories, Berkeley CA 94720}

\author{Michael P. Zaletel}
\affiliation{Station Q, Microsoft Research, Santa Barbara, California, 93106-6105}

\begin{abstract}
{We  determine conditions on the filling of electrons in a crystalline lattice to obtain the equivalent of a band insulator - a gapped insulator with neither symmetry breaking nor fractionalized excitations. We allow for strong  interactions, which precludes a free particle description. Previous approaches that extend the Lieb-Schultz-Mattis argument invoked spin conservation in an essential way, and cannot be applied to the physically interesting case of spin-orbit coupled systems. Here we introduce two approaches, the first an entanglement based scheme, while the second studies the system on an appropriate flat `Bieberbach' manifold to obtain the filling conditions for all 230 space groups. These approaches only assume time reversal rather than spin rotation invariance. The results depend crucially on whether the crystal symmetry is symmorphic. Our results clarify when one may infer the existence of an exotic ground state based on the absence of order, and we point out applications to experimentally realized materials. Extensions to new situations involving purely spin models are also mentioned.}
\end{abstract}

\maketitle
\section{Introduction}
Insulating states of matter arise, in clean systems, as a result of a commensuration between particle density and a crystalline lattice or a magnetic field.  Mott insulators are a particularly interesting class, with an odd number of electrons in each unit cell. Their low energy physics is captured by a spin model with an odd number of $S=1/2$ moments in the unit cell. A powerful result due to Lieb, Schultz, and Mattis in 1D~\cite{Lieb1961}, later extended to higher dimensions by Hastings and Oshikawa~\cite{Oshikawa2000,Hastings2004}, holds that if all symmetries remain unbroken, the ground state must be `exotic' - such as  a Luttinger liquid in 1D, or a quantum spin liquid in higher dimensions, with fractional `spinon' excitations. These exotic states cannot be represented as simple product states, as a consequence of long ranged quantum entanglement. This general result has experimental consequences - indeed no sign of magnetic or spatial symmetry breaking is observed down to temperatures that are orders of magnitude below the intrinsic energy scales in certain materials~\cite{Balents2010}, including the 
quasi 2D Mott insulators $\kappa$-(BEDT-TTF)$_2$Cu$_2$(CN)$_3$, $\beta'$Pd(dmit)$_2$ and Herbertsmithite ZnCu$_3$(OH)$_6$Cl$_2$, as well as the 3D Mott insulator Na$_4$Ir$_3$O$_8$.
Hence if we can apply the Hastings-Oshikawa-LSM theorem (collectively referred to as HOLSM) to these systems, a strong case is made for an exotic ground state (assuming that the effects of disorder can be ignored). However, HOLSM invoke spin rotation invariance in an essential way, which is typically broken in real materials due to spin-orbit coupling. These effects are not small: Herbertsmithite has SO(3)-breaking Dzyaloshinskii-Moriya terms thought to be on the order of 10\% of the Heisenberg coupling~\cite{RigolSingh, Zorko}. In the anti-ferromagnetic hyperkagome compound Na$_4$Ir$_3$O$_8$, the physics is even dominated by spin-orbit coupling effects and charge fluctuation is significant~\cite{Balents2010}.
Physically, the only exact symmetry is time reversal (TR) symmetry, and the crystal translations and charge conservation which allow us to define the filling.
Can HOLSM be extended to this physically more realistic situation?

In this work we show that it indeed can, although entirely different theoretical approaches are needed.
We argue that if a spin-orbit coupled insulator at odd filling is time-reversal symmetric, its ground state must, in a precise sense, be exotic. We introduce two theoretical approaches that, like the flux threading arguments of HOLSM, are non-perturbative, but differs from   them in that conservation of spin is not assumed. The first is an entanglement based approach that allows us to prove that symmetric, gapped and unfractionalized insulators - the interacting analog of a band insulator, with a unique ground state on both the plane and torus - are only allowed at even integer fillings $\nu=2m$.
For brevity we refer to such symmetric short-range entangled insulators as `sym-SRE' insulators.
A corollary of this result is that at odd integer fillings, Mott insulating phases must either break a symmetry or  have a ground state degeneracy on certain geometries due to other, more exotic, mechanisms.
A special case of this result in 1D spin models was previously discussed in Ref.~\onlinecite{Xie2011}.
Here we will extend it to higher dimensions and allow for charge fluctuations. 

This constraint on filling arises even when translations are the only spatial symmetries.
What if additional symmetries are present, such as the 230 space groups of 3D crystals? It turns out that additional constraints appear only for the non-symmorphic space groups, where the minimal filling at which a sym-SRE insulator arises is at least $\nu=4$. We find lower bounds on the minimal filling for all 157 non-symmorphic space groups, and these bounds are shown to be the tightest possible for a large majority of them. Earlier results on noninteracting band structures~\cite{Zak1999,Zak2000,Zak2001} had pointed out that in non-symmorphic crystals there are required band touchings leading to larger minimal fillings. In Refs.~\onlinecite{Sid2013, Roy} this was generalized to interacting systems using flux threading arguments. However, similar to the HOLSM arguments, these need to assume spin rotation invariance, and typically do not lead to useful constraints in their absence. Our entanglement based argument allows both for strong interactions and broken spin rotation invariance, while preserving time reversal symmetry. The best bounds are obtained using a different approach, by defining the system on a closed manifold that is locally flat and hence is locally indistinguishable from Euclidean space (the Bieberbach spaces). These are generalizations of the torus - where instead of identifying points related by translations, one uses non-symmorphic elements, say glides or screws, to obtain a compact manifold. By exposing an obstruction for systems defined on such spaces, we show that the sym-SRE phase, which should be sensitive only to local physics, cannot appear at certain fillings.

Finally we show there are special cases where obstructions to a sym-SRE insulator occurs despite being on symmorphic lattices at even integer filling. These only occur in spin models without charge fluctuations, where reflection symmetry relates two entanglement cuts that can enclose an odd number of sites. 

\section*{Overview}
\label{sec:overview}
Before presenting our detailed arguments, we first give an intuitive description of the key ideas and results.
 
In this work we prove a general constraint on the minimum number of electrons $\nmin$ per primitive unit cell required to realize  a sym-SRE phase:

{\it Consider a system of electrons with Kramers degeneracy. We assume the U(1) (particle number) symmetry, a space group symmetry (including lattice translations), and the time-reversal symmetry. Then, to obtain a sym-SRE phase the average number of electrons per primitive unit cell (i.e.~filling) has to be an even integer no less than two: $\nmin \geq 2$. Furthermore, if the space group is non-symmorphic, $\nmin$ must be an even integer no less than four.}

To warm up, we first ignore charge fluctuations and  consider an infinite, periodic 1D chain with one localized spin-$1/2$ moment per cell. 
For an $S^z$ conserving chain at half filling (including SO(3) invariant chains), the LSM theorem asserts that the ground state must be either gapless or break a symmetry. A more general version of LSM is known in 1D~\cite{Xie2011}: when each unit cell carries a projective representation of a symmetry of the system (or Kramers degeneracy), a sym-SRE phase is forbidden.

\begin{figure}
\begin{centering}
\includegraphics[width=0.5\textwidth]{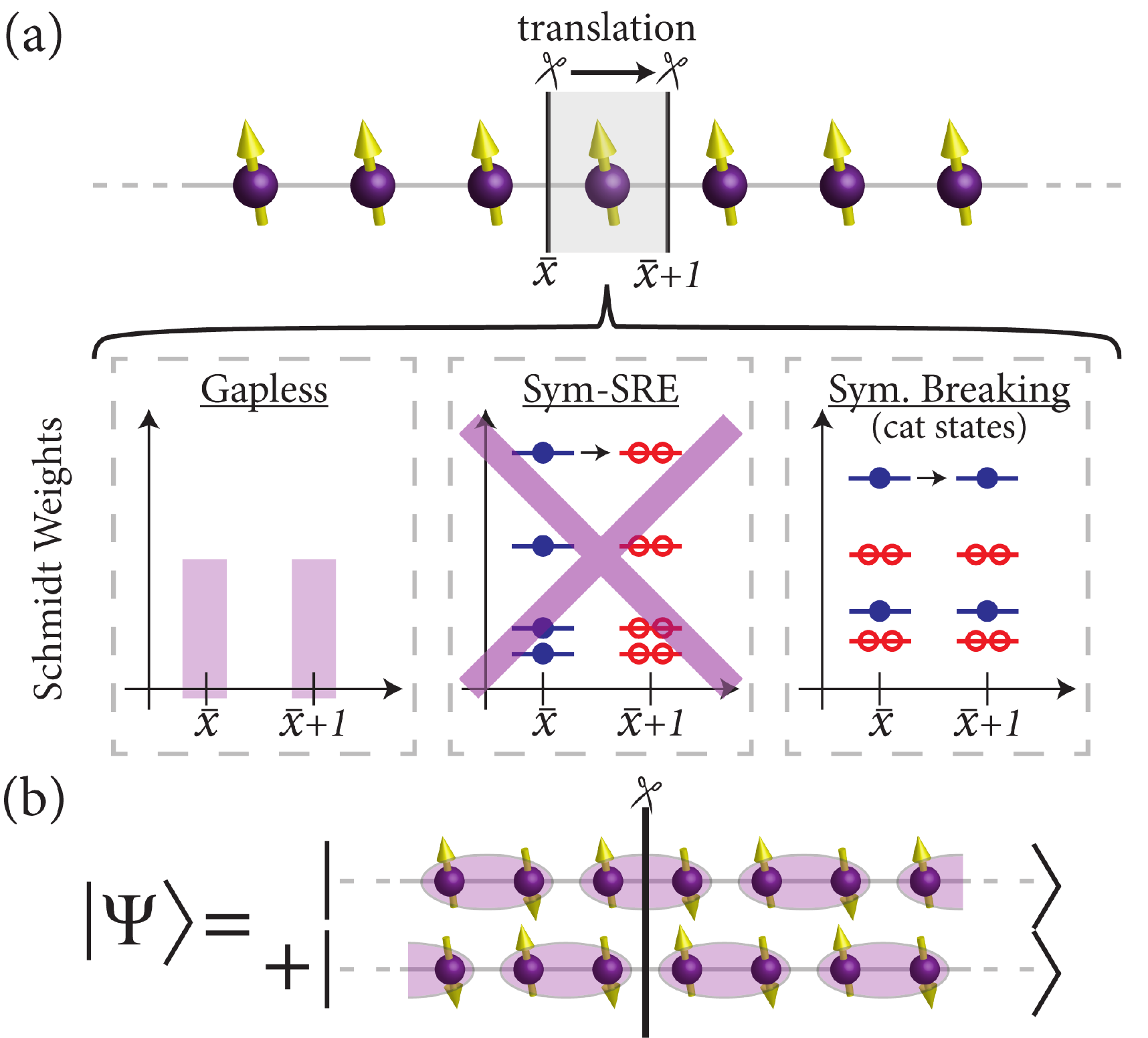}
\caption{
Schematic Schmidt decomposition of an infinite periodic chain with one spin-$1/2$ per cell and time-reversal symmetry. (a) Two entanglement cuts, related by a lattice translation, enclose one spin-$1/2$ moment. If the system is in a sym-SRE phase, the Schmidt states have the same $\hat{\mathcal{T}}^2$ eigenvalue at the same cut and has to change from, say, $\hat{\mathcal{T}}^2=+1$ (blue) at cut $\bar x$ to $\hat{\mathcal{T}}^2=-1$ (red) at cut $\bar x+1$, which is inconsistent with translation symmetry. The system can circumvent such obstruction by either becoming gapless or spontaneously breaking symmetry (such that the symmetric state is a cat state).
(b) Pictorial description of a VBS cat state.\label{fig:SimSchmidt}
}
\end{centering}
\end{figure}

To illustrate the ideas discussed in Ref.~\onlinecite{Xie2011}, consider a spin-$1/2$ chain with TR symmetry $\hat{\mathcal{T}}$.
Suppose the system is in a sym-SRE phase, and consider an entanglement cut at some bond $\bar{x}$, which determines a Schmidt decomposition of the ground state.
Due to the presence of a gap, the high-weight Schmidt states have a discrete spectrum,\cite{Hastings} which we label by $\alpha$. 
The Schmidt states $|\alpha\rangle_L$ are \emph{all} either Kramers singlets, with $\hat{\mathcal{T}}^2 |\alpha\rangle_L = |\alpha\rangle_L$, or doublets, with $\hat{\mathcal{T}}^2 = -1$. This is because in the absence of long-range order, all Schmidt states differ by the action of operators localized near the cut, and local operators never switch the value of $\hat{\mathcal{T}}^2 $.
Now advance the entanglement cut by a lattice translation to bond $\bar{x}+1$, as illustrated in Fig.~\ref{fig:SimSchmidt}(a).
On the one hand, we have added one more spin-$1/2$ to the Schmidt states, so their $\hat{\mathcal{T}}^2$ eigenvalue must change by $-1$; on the other hand, the two cuts should be equivalent due to translation invariance, so their $\hat{\mathcal{T}}^2$ eigenvalue should not change. 
We thus arrive at a contradiction: the Kramers degeneracy carried by each unit cell translates into an obstruction to a sym-SRE phase.

In 1D there are two ways to circumvent the obstruction. If the system is gapless, the Schmidt weights form a degenerate continuum and it is no longer well-defined to track the $\hat{\mathcal{T}}^2$ eigenvalue of any particular Schmidt state.
 If the system spontaneously breaks a discrete symmetry, one can still form a symmetric combination of the ground states, but it will be a Schr{\"o}dinger cat state with long range order.
For a cat state the Schmidt states look different from each other arbitrarily far away from the cut, and the assignment of $\hat{\mathcal{T}}^2$ eigenvalues  may differ across different Schmidt states at the same cut.
For example, for the valence-bond solid (VBS) state in Fig.~\ref{fig:SimSchmidt}(b), we assign ${\mathcal{T}}^2=-1$ if the entanglement cut dissects a singlet pair,  and ${\mathcal{T}}^2=+1$ otherwise. Consequently the $\hat{\mathcal{T}}^2$ eigenvalue depends on the Schmidt state and our argument breaks down.

To obtain similar bounds in higher-dimensional periodic systems, one can extend the above argument by suitably imposing  periodic boundary conditions with odd circumference in all but one dimension to form a quasi-1D cylindrical geometry, then apply the 1D argument.
In dimensions higher than 1D, if the system is gapped the resulting ground state degeneracy at $\nu < \nu_{\text{min}}$ can either be due to spontaneous symmetry breaking, or due to topological order.
For topological orders with $\nu < \nu_{\text{min}}$, the topologically degenerate ground states associated with periodic boundary conditions may be related to each other by the symmetries, which is a subtle form of symmetry-breaking.
For instance, this implies a TR-symmetric system of a Kagome lattice, having three spin-$1/2$ moments per cell, cannot be in the sym-SRE phase.

HOLSM is only one example of a much larger class of constraints we obtain using this program: a sym-SRE phase is actually prohibited whenever two entanglement cuts are related by some spatial symmetry (such as reflection, glide, or screw) and they enclose a projective representation.
In particular, tighter bounds can be obtained when the enclosed volume is only a fraction of the unit cell.
Later in this work we give a concrete model of localized spin moments in which the two cuts are related by a reflection. Consequentially a sym-SRE phase can be forbidden even when there are \emph{two} spin-$1/2$ moments per unit cell.

In real materials, however, the spin-$1/2$ moments typically come from spinful electrons that are not strictly localized and spin-orbit coupling can be significant.
Naively, one might think any lower bound on the required number of localized spin-$1/2$ moments simply morphs into one on electron filling.
Surprisingly, this is not true: there are localized spin models, like the reflection-symmetric one mentioned above, for which a sym-SRE phase ceases to be forbidden once charge fluctuation is allowed.

To understand real materials, it is therefore important to extend HOLSM-type theorems to electronic systems with both charge fluctuation and spin-orbit coupling.
We address this problem in this work assuming only particle number conservation and time-reversal symmetry.
Our results indicate that an electronic system can be in a sym-SRE phase only when the filling is an even integer no less than two.
For \emph{most} non-symmorphic space groups, the presence of fractional translations in the glides or screws allows us to obtain an even tighter bound of $\nu_{\text{min}}\geq 4$.
Two non-symmorphic space groups, however, are exceptional and do not contain any intrinsically non-symmorphic elements~\cite{Mermin};  they instead feature a  `Borromean ring' of screws, and our entanglement argument does not distinguish them from the symmorphic ones.

To handle them, we appeal to the following physical intuition: a sym-SRE phase is sensitive only to local physics, so should have a unique ground state on any large lattice geometry that looks locally identical to the translationally invariant infinite plane. 
These geometries are the lattice analogs of compact flat 3-manifolds, and by systematically studying all the compact flat 3-manifolds we obtain the tightest possible lower bounds for a set of `elementary' space groups that we list in Table~\ref{SummaryTab}. 
Using the group-subgroup relations, we manage to derive lower bounds on the sym-SRE electron filling for each of all 230 space groups (Fig.~2 of Appendix).
These lower bounds are provably tight for all but ten space groups.

\begin{table}
\caption{
Summary of minimal filling $\nmin$ for sym-SRE insulators for `elementary' space groups. \label{SummaryTab}
}
\begin{center}
\begin{tabular}{ccccccc}
\hline
 	& 	& \multicolumn{3}{c}{Minimal Filling}   & (Conway)	
\\
\cline{3-5}
ITC No.		&  	Sym. \tablenote{`Sym.' lists the key symmetry elements of the space groups that are relevant to our discussion. } 	& 	 A.I.\tablenote{The minimal electron filling required to form a TR-symmetric atomic insulator} 	& 	 Ent.\tablenote{Results in section `Extension to 3D symmorphic and non-symmorphic crystals'
} 	 & 	 Bbb.\tablenote{Results in section `Alternative method: Constraining the filling by defining sym-SRE insulators on compact flat manifolds'; for 4 of the elementary NS space groups (No.~19, 24, 29 \& 33) this gives a tighter bound }	 & 	 Manifold name~\cite{Conway2003}	\\
\hline
{\bf \emph 1}\tablenote{For completeness we include results for the `elementary' symmorphic space group}	& (translation)	&2&2&2&Torus $T^3$\\
\hline
{\bf \emph 4}					&$2_1$ 		 	&4&4&4&Dicosm\\
\hline
{\bf \emph 144}/{\bf \emph 145}		&$3_1$/$3_2$   	&6&6&6&Tricosm\\
\hline
 {\bf \emph 76}/{\bf \emph 78} 		&  $4_1$/$4_3$ 	&8&8&8&Tetracosm\\
 {\bf \emph 77} 		 			& $4_2$ 			&4&4&4&	\\
 {\bf \emph 80} 		 			& $4_1$+BCT\tablenote{BCT: Body-centered translation; the minimal filling is therefore half of the corresponding primitive space group} 	&4&4&4&\\
\hline
 {\bf \emph 169}/{\bf \emph 170} 	& $6_1$/$6_5$ 	&12 &12&12&Hexacosm\\
 {\bf \emph 171}/{\bf \emph 172} 	& $6_2$/$6_4$ 	&6&6&6&\\
 {\bf \emph 173} 		 		& $6_3$ 			&4&4&4&\\
\hline
{\bf \emph 19}					& $2_1 2_1$ 		&8&4&8&Didicosm\\
{\bf \emph 24} 					& $2_1 2_1$+BCT  	&4&2&4&\\
\hline
 {\bf \emph 7}					& glide   	 		&4&4&4&$1^{\text{st}}$ Amphicosm\\
\hline
 {\bf \emph 9} 					& glide   			&4&4&4&$2^{\text{nd}}$ Amphicosm\\
\hline
 {\bf \emph 29}					& glide+$2_1$  	 	&8&4&8& $1^{\text{st}}$ Amphidicosm\\
\hline
{\bf \emph 33}					& glide+$2_1$		&8&4&8&$2^{\text{nd}}$ Amphidicosm\\
\hline
\end{tabular}
\end{center}
\end{table}

\subsection*{A LSM-type theorem for interacting spin-orbit coupled chains}
\label{sec:SOC_LSM}
Here we prove that a TR-invariant 1D interacting SOC chain at odd filling $\nu$ is either gapless or breaks symmetry (translation or TR).  The argument presented here paraphrases a more precise proof using the language of matrix product states in Appendix.
We proceed by assuming the ground state $\ket{\Psi}$ is symmetric and short-range correlated, and derive a constraint on the filling $\nu$.
Let $x$ label unit cells and $\hat T_1$ be the translation $x \to x+1$.
Each unit cell may contain many sites and orbitals.
We Schmidt decompose $\ket{\Psi}$ across a bond ${\bar{x} \in ( x_0-1, x_0)}$,
\begin{align}
\ket{\Psi} = \sum_{\alpha} s_{\bar{x} \alpha} \ket{\alpha}_{\bar{x}\, L} \ket{\alpha}_{\bar{x}\, R}
\end{align}
with Schmidt weights $s_{\bar{x} \alpha}$. 
Assuming the phase is gapped, $\ket{\Psi}$ obeys an area law, so the spectrum $s_{\bar{x} \alpha}$ (and hence the label $\alpha$) is discrete.
Each Schmidt state is an eigenstate of the number operator, but it is subtle to discuss their eigenvalues (the total charge to the left) as they are ill-defined in the thermodynamic limit.  Instead, the well-defined quantity is the charge relative to the mean filling, $\hat{Q}_x = \hat{N}_x - \nu$.
We have
\begin{align}
\sum_{x < \bar{x}} \hat{Q}_x  \ket{\alpha}_{\bar{x}\, L}  = Q_\alpha  \ket{\alpha}_{\bar{x}\, L},\,\,\, Q_\alpha = \sum_{x < \bar{x}}  \bra{\alpha}  \hat{Q}_x \ket{\alpha}_{\bar{x}\, L}.
\end{align}
The sum is well conditioned because $\bra{\alpha}  \hat{Q}_x \ket{\alpha}_{\bar{x}\, L} \to 0$ as $x \to -\infty$ exponentially quickly for a gapped state. This follows from the fact that the Schmidt states differ from the ground state only in the vicinity of the cut and the ground state is assumed not to have charge-density wave (CDW) order.

Translation invariance $\hat{T}_1\ket{\Psi}=\ket{\Psi}$ implies that the Schmidt decompositions across different bonds are related by $s_{\bar{x} \alpha} = s_{\bar{x} + 1\,  \alpha}$, $\hat{T}_1 \ket{\alpha}_{\bar{x} \, L/R} =  \ket{\alpha}_{\bar{x} + 1 \, L/R}$, with the two translation-related states having the same charge excess $Q_\alpha$. 
The Schmidt states at the two cuts are related by the addition of the intervening sites, spanned by states $\ket{p}_{x}$:
\begin{align}
\ket{\alpha}_{\bar{x} + 1 \, L} = \sum_{p, \beta} B^p_{\alpha \beta} \ket{p}_{x} \ket{\beta}_{\bar{x}\, L}.
\label{decomposition}
\end{align}
Choosing a charge eigenbasis for the sites, $\hat{Q}_{x} \ket{p}_{x} = Q_p \ket{p}_{x}$ with $Q_p\in \mathbb{Z}-\nu$, charge conservation implies that whenever $B^p_{\alpha \beta} \neq 0$ we have $Q_\alpha = Q_\beta + Q_p$.
However, so long as the state is short-range correlated, $Q_\alpha-Q_\beta \in \mathbb{Z}$, since any two Schmidt states differ by the addition or rearrangment of some particles near the entanglement cut.
This is a contradiction unless $\nu\in\mathbb{Z}$ (the HOLSM theorem).  
If $\ket{\Psi}$ is a cat state of two CDWs, so is symmetric but not short-range correlated,  the differences $Q_\alpha-Q_\beta$ no longer need to be an integer and our argument does not apply.

When TR symmetry is incorporated, each Schmidt state is either part of a Kramers singlet or doublet:
\begin{align}
\hat{\mathcal{T}}^2 \ket{\alpha}_{\bar{x}\, L} = (\mathcal{T}^2)_\alpha  \ket{\alpha}_{\bar{x}\, L}, \quad (\mathcal{T}^2)_\alpha =  \pm1.
\end{align}
Of course $(\mathcal{T}^2)_\alpha$ is related to $(-1)^{Q_\alpha}$, but we have to recall that $\hat{\mathcal{T}}^2$ is defined to be $(-1)^{\hat{N}_{x}}$, not $(-1)^{\hat{Q}_{x}}$. 
The divergence of the total charge results in an ambiguous but (assuming short-range correlations) $\alpha$-independent phase:
\begin{align}
(\mathcal{T}^2)_\alpha = e^{i \Phi} (-1)^{Q_\alpha}.
\label{T2eig}
\end{align}
Again, because of translation invariance, the charge excess $Q_\alpha$ and the TR character $(\mathcal{T}^2)_\alpha$ are independent of $\bar{x}$. 
 
Now consider the action of $\hat{\mathcal{T}}^2$ on the decomposition \eqref{decomposition}:
\begin{align}
\hat{\mathcal{T}}^2 \ket{\alpha}_{\bar{x} + 1 \, L} &= \sum_{p, \beta} B^p_{\alpha \beta} (\hat{\mathcal{T}}^2 \ket{p}_{x_0})(\hat{ \mathcal{T}}^2 \ket{\beta}_{\bar{x}\, L})\notag\\
&=  e^{i \Phi} \sum_{p, \beta} B^p_{\alpha \beta}  (-1)^{Q_p + \nu + Q_\beta}  \ket{p}_{x_0} \ket{\beta}_{\bar{x}\, L}\notag\\
&=  (-1)^{\nu} e^{i \Phi} (-1)^{Q_\alpha}\ket{\alpha}_{\bar{x}+1 \, L},
\end{align}
where we used $Q_\alpha = Q_\beta + Q_p$ and Eq.~\eqref{decomposition} again in the last step.  Note the important factor of $(-1)^{\nu}$ compared to $(\mathcal{T}^2)_\alpha$ in Eq.~\eqref{T2eig}.  It follows that $(-1)^\nu=1$ and hence $\nu \in 2 \mathbb{Z}$.  

If $\nu \notin 2 \mathbb{Z}$, one of our assumptions must be violated: either the phase is gapless, the state breaks a symmetry, or it is symmetric but a long-range correlated cat state.
For example, for a gapless system the entanglement Hamiltonian is also gapless: all $s_\alpha \to 0$, the Schmidt states form a continuum, and the symmetry properties of the Schmidt states are ill-defined.
Alternatively, for a VBS cat state in Fig.~\ref{fig:SimSchmidt}(b), the ambiguous phase $e^{i \Phi}$ is no longer $\alpha$-independent, but instead will depend on which of the two VBS sectors the Schmidt state belongs to.

\subsection{Extension to 3D symmorphic and non-symmorphic crystals}
\label{sec:Extension}
The above argument can be extended to 3D systems
\footnote{Our results in this and the next sections are also applicable to 2D systems characterized by layer groups, since any layer group has a natural correspondence to a space group.} 
using the quasi-1D (generalized) cylinder $\mathbb{R} \times T^2$.  
Given a 3D crystal with the primitive lattice vector $\vec{a}_{1,2,3}$, we introduce a periodic boundary condition that identifies $\vec{x}$ with $\vec{x}+ N_2 \vec{a}_2$ and $\vec x+ N_3 \vec{a}_3$.
We are then left with a single `long' direction with infinite extent along $\vec{a}_1$ and a translation symmetry $\hat T_1$.
Formally we can view the cylinder as a 1D-chain with a 1D-unit cell that comprises $N_2N_3$ of the 3D unit cells.
Therefore, choosing $N_2, N_3$ to be odd, the constraint on 1D filling $\nu_{\oned} = (N_2 N_3) \nu\in 2 \mathbb{Z}$ implies $\nu \in 2 \mathbb{Z}$.

The extension implicitly assumed the following: if the phase is trivial, for sufficiently large $N_2, N_3$ the resulting cylinder must also have a unique ground state.
Intuitively, a sym-SRE phase cannot detect the global topology - we will comment further on this later.
Similar assumptions were invoked in Oshikawa's argument~\cite{Oshikawa2000}.

The lower bound $\nmin=2$ is in fact the tightest bound for all $73$ \emph{symmorphic} space groups in 3D (unless extra constraints, like specifying a lattice realization, are imposed).
A space group is said to be symmorphic if every symmetry element can be decomposed into a product of a lattice translation $n_i\vec{a}_i$ ($n_i\in\mathbb{Z}$) and a point group element.  
Such a decomposition implies that putting a site at the `origin' of the point group within each unit cell generates a lattice that respects all the symmetries.
For symmorphic space groups, therefore, one can trivially realize an {\em atomic insulator} at $\nu=2$ by putting a pair of electrons into a singlet state at the highest symmetry point in each unit cell.  Since we can explicitly construct a sym-SRE state at $\nu = 2$, the bound $\nmin=2$ is tight.
Note that such a high-symmetry point is a special instance of the allowed positions of sites compatible with the space group, known as the {\em `Wyckoff positions'}.
A set of sites in a Wyckoff position (labeled by a `Wyckoff letter') transforms within itself under all symmetry elements, and can be thought of as the admissible positions of the atoms. 

The situation is very different when the space group is non-symmorphic (NS), for which given any fixed origin at least one symmetry element contains a fractional translation. 
As a direct consequence, each unit cell must contain more than one site.
Since an atomic insulator is formed by having a singlet pair of electrons per \emph{site}, the \emph{upper} bound on $\nu_{\text{min}}$ per cell can be inferred from the Wyckoff position with fewest sites.  This is listed in the `A.I.' column of Table~\ref{SummaryTab}.

To systematically study all $157$ NS space groups, we focus on those $18$ space groups listed in Table~\ref{SummaryTab}, which we refer to as `elementary' in the following.  All other NS space groups contain at least one of these elementary groups as a subgroup. (More precisely, here we are referring to so-called {\it translationengleiche} subgroups or $t$-subgroups~\cite{ITC}, which retain the primitive unit cell of the entire group.) Thus, a lower bound on the elementary space groups immediately provides a lower bound for the others, though further work is required to determine whether the inferred bounds are tight.

All elementary NS space groups, except for No.~24, are made NS by a $n_m$-screw ($2\pi/n$-rotation $\hat{C}_n$ followed by $m/n$-translation) or a glide (a mirror reflection followed by a half translation), which we denote by $\hat{G}$ and list in the `Sym.~elements' columm of Table~\ref{SummaryTab}.  
No.~24 is known as an `exceptional' NS space group and will be addressed later.

To establish a dimensional reduction as before, we have to specify our choice of the primitive lattice vectors $\vec{a}_{1,2,3}$.
We require the following: (i) the NS operation $\hat{G}$ is represented as $\hat{G}=e^{-i(m/n)\hat{\vec{P}}\cdot\vec{a}_1} \hat{X}$.
Here, $\hat{X}$ is the rotation $\hat{C}_n$ about an axis parallel to $\vec{a}_1$ for $n_m$-screws, or the mirror about a plane that contains $\vec{a}_1$ for a glide (where $m/n=1/2$). (ii) The plane spanned by $\vec{a}_2$ and $\vec{a}_3$ is invariant under $\hat{X}$.
For these elementary NS groups, one can check case by case that such a choice is possible. (For No.~80, one has to use the conventional cell, which contains two primitive unit cells.)

\begin{figure}
\begin{centering}
	\includegraphics[width=0.5\textwidth]{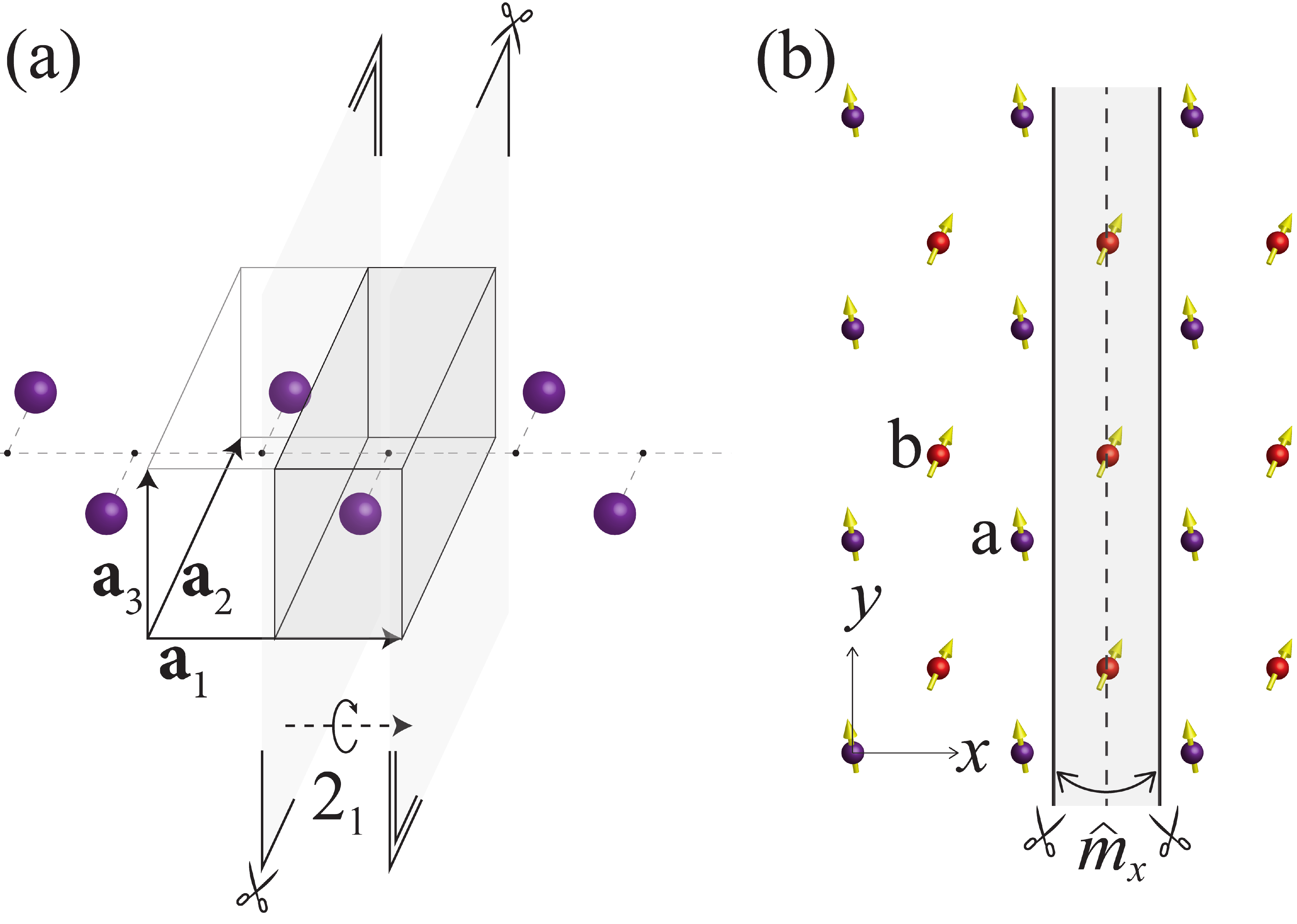}
	\caption{Entanglement cuts enclosing fractions of the unit cell. (a) In the quasi-1D setup,  two entanglement cuts  related by $\hat{G}$, say a $2_1$ screw,  enclose a half-integral volume.
	(b) Distorted checkerboard lattice. Two entanglement cuts are related by the reflection symmetry.\label{fig:EntCuts}}
\end{centering}
\end{figure}

As before, we take a cylinder geometry $\mathbb{R}\times T^2$ by introducing the periodic boundary condition for $N_2\vec{a}_2$ and $N_3\vec{a}_3$. 
We then replace the translation $\hat{T}_1$ in the above argument with the NS operation $\hat{G}$.
The action of $\hat{X}$ is purely `onsite' in the 1D picture and has no effect on our 1D argument, since $\hat{X}$ merely rotates or reflects the plane spanned by $\vec{a}_2$ and $\vec{a}_3$, as illustrated in Fig.~\ref{fig:EntCuts}(a).  
The volume enclosed by the two cuts related by $\hat{G}$ is $m/n$ times smaller than the symmorphic case, so we now have $\nu_{\oned} =  (m/n) N_2 N_3\nu$.  
Requiring $\nu_{\oned} \in 2 \mathbb{Z}$, we prove the $\nu_{\text{min}}$ listed under column `Ent.' in Table~\ref{SummaryTab}.  (Remember that the conventional cell of No.~80 contains two primitive unit cells, so we have to divide the naive bound $8$ by two to get the filling bound of $4$.)  For most of the elementary NS space groups, the lower bound obtained here coincides with the upper bound from the atomic insulator limit, and hence is also the tightest bound.

Curiously, the argument here does not produce a tighter bound for No.~24. The difficulty is that No.~24 contains \emph{three} orthogonal $2_1$ screws, and if any one is absent the crystal is symmorphic.  Therefore, we need a new machinery that makes use of multiple NS operations at once. Similarly, the bounds obtained for No.~19,~29 and 33 are not the tightest because they are also generated by multiple NS operations.


\subsection*{Alternative method: Constraining the filling by defining sym-SRE insulators on compact flat manifolds }
\label{sec:bieberbach}
The above argument based on the 1D Schmidt decomposition can only utilize a single glide or screw, whereas multiple such non-symmorphic operations can be present in a space group.
Here we present a different line of argument to overcome this challenge.  

We employ the following physical principle: if a Hamiltonian defined on an infinite lattice $\Lambda$ with space group $\SG$ is in a sym-SRE phase, it will have a unique ground state on any compact lattice geometry in which 
a) the Hamiltonian is locally indistinguishable from that on $\Lambda$,
b) the particle density is that of the thermodynamic limit, and 
c) the width of all compact directions is sufficiently larger than some scale comparable to the correlation length. 
The lattice geometries we have in mind are  generalizations of the torus; in the continuum we would say they are `flat'.
Intuitively, a sym-SRE state is only sensitive to local physics, so cannot detect the global geometry so long as each patch is locally identical to $\Lambda$.

Let us first use this principle to reproduce our constraint on $\nu$ for symmorphic crystals.
To that end, it is sufficient to put the system on a torus $T^3$ with odd circumferences $N_1, N_2, N_3$ in every direction, which is a well defined procedure since the Hamiltonian is translationally invariant.
If the filling $\nu$ is odd, the total number of electrons $N_e = \nu N_1 N_2 N_3$ is also odd and Kramers degeneracy is unavoidable.
Yet according to the principle, a sym-SRE phase should have a unique ground state: it is short-range correlated, so cannot detect the parity of $N_i$ once $N_i \gg 1$.
Hence, to realize a sym-SRE phase, $\nu$ must be even.

The advantage of this argument is that one may get a stronger constraint on $\nu$ by using manifolds other that the torus.
To this end, we discuss how to define a lattice model on a flat manifold obtained by `modding-out' $\mathbb{R}^3$ by a subgroup $\Gamma$ of the $\SG$.
Let us first formalize the familiar example of putting a translation invariant 1D system onto a ring $S^1$.
If the circumference of the ring is $N$, we identify sites $x$ with $x+n N$ ($n\in\mathbb{Z}$) and operators $\hat{O}_x$ with $\hat{O}_{x+nN}$.
Denoting $\Gamma$ as the group generated by an $N$-step translation, one can formally rewrite this identification rule ($\sim_\Gamma$) as
\begin{eqnarray}
\vec{x}\sim_\Gamma g(\vec{x}),\quad \hat{O}_{\vec{x}}^{(i)}\sim_\Gamma \hat{g}\hat{O}_{\vec{x}}^{(i)}\hat{g}^{-1}=\hat{O}_{g(\vec{x})}^{(j)} (U_g)_{ji}
\label{equivalence}
\end{eqnarray}
for $g \in \Gamma$. Here, $\hat{g}$ is the quantum operator corresponding to $g$, $\hat{O}_{\vec{x}}^{(i)}$ runs over all the operators at $\vec{x}$, and $U_g$ is a unitary matrix representation which could account for any action of $g$ on the internal degrees of freedom, as will eventually be required for spin-orbit coupling.
For the usual notion of translation, $U_g = 1$.
The Hamiltonian for the ring is obtained from that of the infinite chain by identifying terms using the equivalence $\sim_\Gamma$.

The same procedure applies to more complicated space group symmetries: given a 3D lattice Hamiltonian symmetric under $\Gamma \subset \textrm{SG}$, we impose the equivalence relation $\sim_\Gamma$ in order to obtain a lattice model on the space $\mathbb R^3 / \Gamma$.
There are two requirements on $\Gamma$ in order to obtain a well-defined lattice Hamiltonian.
First, $\Gamma$ must be fixed-point free (also called 'Bieberbach'), as otherwise the manifold will contain conical singularities or edges, so won't be locally indistinguishable from the plane (see Appendix for more details).
Second, since $U_g$ also encodes the nontrivial action of glides and screws on the internal spin, $\Gamma$ has to be compatible with certain consistency conditions (namely the existence of the lattice analog of Spin and Pin$^-$ structure), which we will return to shortly.

There are 10 compact flat manifolds (also known as Bieberbach manifolds) in 3D, and each of them can be obtained by modding-out $\mathbb{R}^3$ by a Bieberbach group $\Gamma$~\cite{Conway2003}.
For example, when $\Gamma$ is generated by a glide and two other orthogonal translations we obtain a manifold diffeomorphic to $(\text{the Klein bottle})\times S^1$, the so-called `first amphicosm.'~\cite{Conway2003}
When $\Gamma$ is non-symmorphic, the number of unit cells contained in the resulting manifold $\mathbb{R}^3/\Gamma$ can be \emph{fractional}, and will lead to a tighter bound.

\begin{figure}
\begin{center}
{\includegraphics[width=0.5\textwidth]{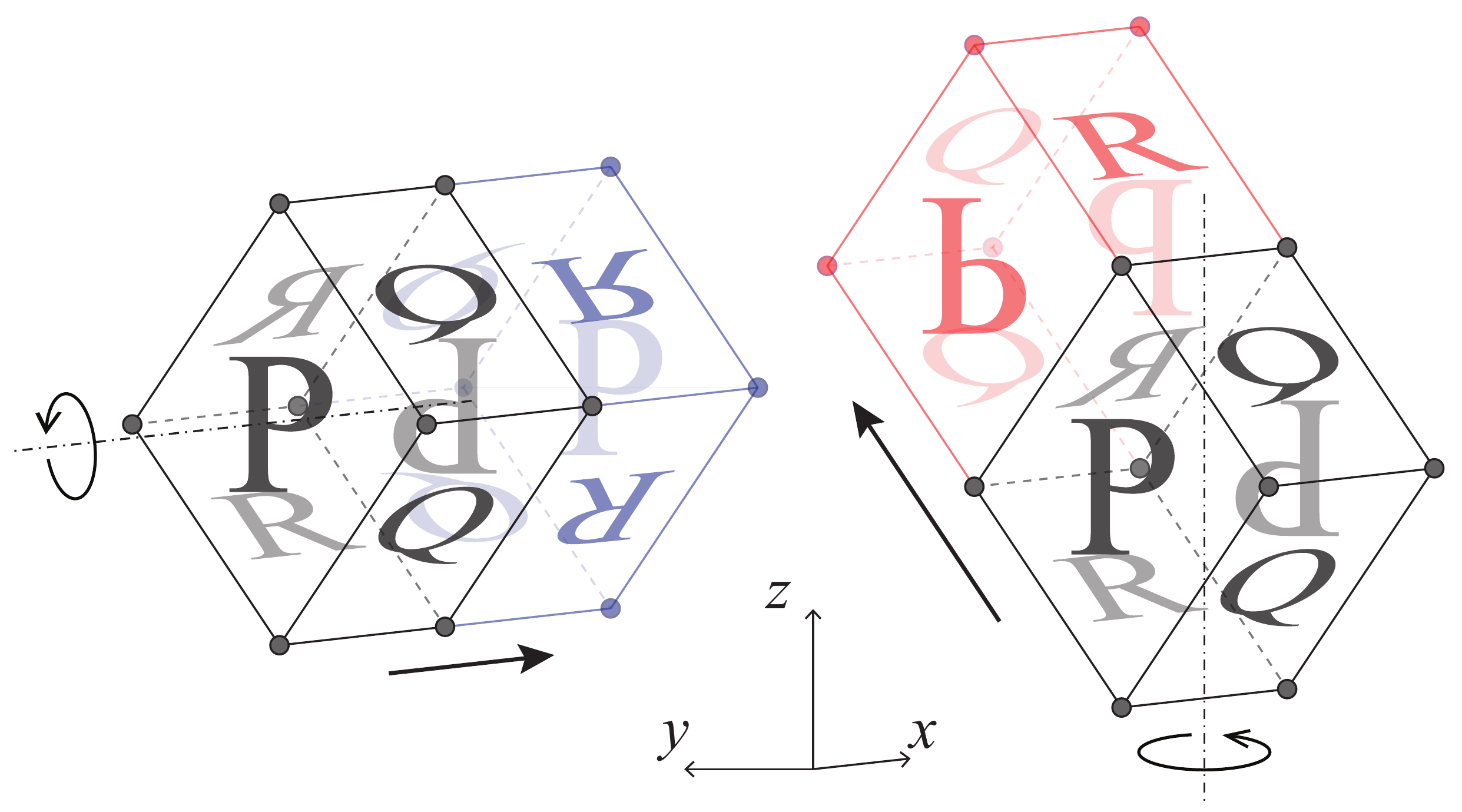}} 
\caption{\label{didicosm} Illustration of a Didicosm obtained by modding out $\mathbb{R}^3$ by a suitably chosen subgroup of space group No.~24. This space group is generated by two orthogonal screws.  The black box represents a fundamental domain, and the blue and red boxes are the image of the fundamental domain under the two screws. 
The letters P, Q and R and their orientation indicate pairs of faces that are identified.
}
\end{center}
\end{figure}

For example, to derive the constraint for space group No.~24, 
we put the system on a manifold obtained from a subgroup $\Gamma$ spanned by two suitably chosen screws.
The resulting manifold (Fig.~\ref{didicosm}) can contain a half-integer number of primitive unit cells of No.~24. 
Specifically, the number of electrons will be $N_e = \nu N_1 N_2 (2 N_3+1)/ 2$ for $N_i \in \mathbb{Z}$. Choosing $N_1 N_2$ to be odd,  we must have $\nu\in 4 \mathbb{Z}$ to avoid a Kramers degeneracy.

By engineering $\Gamma$ for each of the elementary space groups, we have obtained the tight bounds $\nmin$, as summarized in Table~\ref{SummaryTab}.  Furthermore, using group-subgroup relations described in detail in Appendix, we determine the set of all possible fillings that are allowed to realize a sym-SRE phase for each of all 157 NS space groups, which we include in Appendix. 
Except for 10 NS space groups (No.~73, 106, 110, 133, 135, 142, 206, 220, 228 and 230), these fillings are necessary and sufficient - there is indeed a noninteracting band insulator at these filling~\cite{us2}.
For the 10 NS groups mentioned, however, our method here indicates a sym-SRE insulator is possible at $\nu =4$, whereas one can in fact show a band insulator at the same filling is impossible~\cite{us3}. 
Therefore, there are two possibilities --- either one can actually show a stronger constraint that prohibits this filling (by utilizing point group symmetries of the space group, which we did not use in this work) or there indeed exists an interacting example of sym-SRE phase at this filling. We leave this open problem to future work.

In closing we note the following technical point.
It is well known that consistently defining spinors in the continuum requires the manifold to admit `Spin' or `$\text{Pin}^\pm$' structure, so we should investigate whether this constraint arises in the SOC lattice setting ($\text{Pin}^\pm$ structures extend Spin structures to non-orientable manifolds; the $\pm$ sign indicates that the square of a reflection is $\pm1$).
The subtlety arises because SOC spinful fermion operators $\hat{f}_{\vec{x} \, \sigma}$ transform under a \emph{double} group of the $\SG$ [e.g., SU(2) rather than SO(3) for the proper rotation part of the $\SG$], leading to sign ambiguities in the choice of $U_g$ in Eq.~\eqref{equivalence}.
If the signs are not fixed properly, the $U_g$ will generate a projective representation of $\Gamma$, i.e., $U_{g_1}U_{g_2}=\pm U_{g_1g_2}$.
When the $U_g$ are projective, the equivalence relation $\sim_\Gamma$ is ill-defined; we do not know if we should identify $U_{g_1g_2}\hat{f}_{g_1g_2(\vec{x})}$ with $+\hat{f}_{\vec{x}}$ or $-\hat{f}_{\vec{x}}$. Therefore, we have to carefully choose the sign of each $U_g$ in such a way that $U_g$ forms a linear (nonprojective) representation of $\Gamma$.
This choice may not be unique, and as detailed in Appendix each choice corresponds to a different Spin / Pin$^-$ structure~\cite{Lutowski2014}.
Fortunately, all 3D flat manifolds admit either a Spin or Pin$^-$ structure, so the proper choice of $U_g$ is always possible.


\subsection*{Localized spin models: New symmetry constraints}
\label{sec:beyondB}
So far we have focused on universal constraints on $\nu$ with the input of space groups only. 
We now show that, given specific positions of sites in the unit cell, there is an entirely new class of constraints.
Unlike the previous section where we could utilize only fixed-point free symmetry operations, 
here we use a point-group symmetry to probe the \emph{location} of the sites in the unit cell.  For this result we must restrict to the strongly Mott-insulating regime where the system reduces to $S=1/2$ moments at lattices sites.
In the presence of certain reflection symmetries, even for symmorphic crystal lattices the minimal filling for a trivial paramagnet can be $\nmin= 4$, not $\nmin = 2$.  In particular: \emph{If an $S=1/2$ magnet has an odd number of moments per unit cell lying on a reflection plane, the system is either gapless, or the ground state is not-unique.}  The `tighest' bound is shattered because here we assume a more constraining scenario, namely we assume specific lattice realizations of the space group symmetries where the electrons are strongly localized.

As a concrete example we consider the distorted checkerboard lattice shown in Fig.~\ref{fig:EntCuts}(b) with $S=1/2$ moments localized on every lattice site.
This is an example of a lattice with the symmorphic wallpaper group symmetry No.~3 ($pm$), which contains a mirror $(x,y) \rightarrow (-x,y)$. There are two inequivalent mirror planes $x=0$ and $x=1/2$, and we consider a lattice formed by putting one site on each of the two planes in each unit cell as shown in Fig.~\ref{fig:EntCuts}(b).
No symmetry relates the moments on the two sub-lattices $a$ and $b$ (corresponding to Wyckoff letters $a$ and $b$ in Ref.~\onlinecite{ITC}), and the crystal is symmorphic, so our previous results do not imply a constraint. 

Now we consider putting the system onto a cylinder with an odd circumference in the $y$ direction. Consider an entanglement cut at anywhere between $x=0$ and $1/2$. Under mirror reflection about the plane $x=1/2$, the two entanglement cuts enclose an odd number of sites in sublattice $b$ and hence an odd number of $S=1/2$ moments. Now suppose that each $S=1/2$ moment carries a projective representation of some symmetry, say TR or SO(3).  Then these symmetry-related entanglement cuts yield corresponding Schmidt states carrying different representations. This rules out the possibility of having a trivial ground state at filling $\nu=2$. 

Note, however, if charge fluctuations are allowed, then we can localize both electrons to one or the other lattice site and thereby obtain a sym-SRE insulator. Therefore it was important that we consider the pure spin model. Another route to preserving the obstruction is to consider Hamiltonians with particle-hole symmetry, obtained for instance, by only allowing hopping between bipartite lattice sites. This also effectively constrains the filling on sites.


\section*{Applications and Conclusion}
\label{sec:conclusion}
Let us discuss some physical applications of our result. First, consider the statement - a Mott insulator with an odd number of electrons per unit cell must either be gapless or, if gapped, must display a ground state degeneracy. This is true even when only time reversal symmetry is present and  spin rotation invariance is absent. The Mott insulator has a net Kramers doublet per unit cell, which is the analog of having a half odd integer spin per unit cell.  This is of practical relevance in real materials, where spin-orbit coupling always breaks the full spin rotation symmetry while time reversal continues to be a physical symmetry. Let us consider the possible ground states of a Mott insulator with just time reversal symmetry in $D=2$. The conventional ground states either spontaneously break time reversal symmetry, or spatial translation symmetry. In both cases ground state degeneracies result due to symmetry breaking. However, any other cause for ground state degeneracy, or gaplessness, must have an `exotic' origin. For example, ground state degeneracy that is not attributable to symmetry breaking implies there is no local operator that could distinguish the different states. This is the hallmark of topological order, as in a  quantum spin liquid, where emergent excitations with self or mutual anyonic statistics are expected. Similarly, the conventional origin of gapless excitations in a phase, either a Fermi sea of electrons or Goldstone modes are both ruled out since we are dealing with a symmetric insulator. The materials Herbertsmithite (ZnCu$_3$(OH)$_6$Cl$_2$) is a Mott insulator with spin $1/2$ moments on a Kagome lattice~\cite{Helton}. The ground state is believed to be symmetric since no thermodynamic phase transition occurs on lowering the temperature to a small fraction of the exchange energy. In addition to Heisenberg exchange terms, spin-orbit coupling induced Dyaloshinskii-Moriya interactions lower the spin rotation symmetry~\cite{Zorko}. Hence one cannot directly apply the Oshikawa-Hastings type argument to this system. However, since time reversal is preserved, we can use the results of the present paper to conclude that the resulting ground state must be an unconventional phase of matter such as a gapped quantum spin liquid. 

A more surprising application of our results concerns the truly three-dimensional material Na$_4$Ir$_3$O$_8$, in which the iridium atoms form a hyperkagome lattice with 12 sites per primitive unit cell~\cite{Takagi}. Similar to Herbertshimite, no thermodynamic phase transition is observed when approaching the zero-temperature limit and the ground state is believed to be symmetric.
In a simplified picture, the system can be described as an anti-ferromagnetic Heisenberg model with one spin-1/2 moment on each of the iridium sites. In reality, however, both spin-orbit coupling and charge fluctuation are significant in Na$_4$Ir$_3$O$_8$~\cite{Balents2010}. This precludes any application of the original HOLSM theorem. While our results have taken into account such generalizations, the electron filling here is even and so the argument above, utilizing the odd electron filling in Herberstmithite, is also inapplicable. Nonetheless, we can still rule out a sym-SRE insulating phase by incorporating the non-symmorphic space group symmetries:
Na$_4$Ir$_3$O$_8$ crystallizes in the space group No.~213 ($P4_1 32$)~\cite{Takagi}, which admits a sym-SRE insulator only when $\nu \in 8 \mathbb Z$ (Appendix Table I) but $\nu = 12$.

One caveat though is the presence of impurities strictly speaking, destroys translation invariance which is crucial to defining filling~\cite{Helton}. 
In addition, for some systems certain atoms are not perfectly ordered and the space group symmetries are, strictly speaking, only approximate. This is the case in the hyperkagome Na$_4$Ir$_3$O$_8$, in which the sodium atoms sit at Wyckoff positions $4a$ and $12d$ with only $75\%$ occupation~\cite{Takagi}.
Hence all statements are about a putative clean system in which the relevant degrees of freedom are symmetric under the space group considered.

As a second example consider the Dirac semimetal described in Refs.~\onlinecite{Young2012,Young2015}. A simple model of this state is the Fu-Kane-Mele~\cite{FuKaneMele} model of electrons on a diamond lattice at half filling, which leads to Dirac cones at face centers of the Brillouin zone.  In a noninteracting electron picture, the stability of this interesting electronic structure arises from the non-symmorphic space group of the diamond lattice, which leads to a four dimensional irreducible representation which forms the apex of the Dirac cone. Can strong interactions gap these nodes and lead to a sym-SRE insulator? Since the diamond lattice has two sites in the unit cell, the filling of two electrons is compatible with a sym-SRE state in a symmorphic crystal. However, given the non-symmorphic diamond lattice, the minimum trivial filling is four - implying that if the Dirac nodes are gapped the system necessarily breaks symmetry or is topologically ordered.  

In conclusion, we have obtained the allowed sequences of filling for which SOC-coupled interacting electrons can form a sym-SRE insulator for all elementary space groups. For symmorphic lattices this happens to be an even number of electrons per unit cell, while for non-symmorphic lattices the sequence is more restrictive and demands at least four electrons per unit cell. The minimal fillings for a variety of non-symmorphic space groups were derived, however a complete exploration of the tightness of these lower bounds is left for future work. Similarly, the exploration of constraints on  topological order and symmetry fractionalization for non-symmorphic lattices, analogous to the analysis for Mott insulators~\cite{MZAV}, is an interesting open problem.

\begin{acknowledgments}
MZ is indebted to conversations with Mike Freedman, Matt Hastings, Mike Hermele,  and Meng Cheng and thanks Sid Parameswaran for the introduction to NS space groups. AV thanks S. Parameswan, Ari Turner and Dan Arovas for an earlier collaboration on related topics, Leon Balents for insightful discussions and support from NSF DMR 1206728 and the Templeton Foundation.  HW thanks Takahiro Morimoto, Yohei Fuji, Masaki Oshikawa, and Hitoshi Murayama for useful comments.
HCP was supported by a Hellman graduate fellowship and   NSF DMR 1206728.
\end{acknowledgments}

\bibliography{references}

\appendix

\section{Proof of minimal spin-orbit coupled filling in the matrix-product-state formalism.}
\label{sec:aMPS}
In the MPS formalism, a short-range correlated translation invariant wave-function $\Psi$  is written as
\begin{align}
\Psi_{ \{ a_n \} } =  \sum_{ \{\alpha_n\} = 1}^\chi  \prod_n B^{a_n}_{\alpha_n \alpha_{n+1}},
\end{align}
where the local Hilbert spaces are indexed by $a_n$ and the $\alpha_n$ are ``auxiliary'' indices to be summed over. 
For our purposes, $\chi$ can actually be infinite: what's important is that the Schmidt decomposition has a well behaved and discrete spectrum, which follows from the area-law obeyed by gapped ground states~\cite{Hastings}.
An important question is how the global symmetries of $\Psi$ manifest themselves in the local tensors $B$.
In our case the global U(1) and time reversal symmetries factorize into their onsite representations,
\begin{align}
\hat{G}_\theta = \prod_n e^{i \theta N_n} = \prod_n g^{(n)}_\theta,\\
\hat{G}_{\mathcal{T}} K =  \prod_n  g^{(n)}_{\mathcal{T}} K,
\end{align}	
where $n$ label sites and the $g$ are onsite unitary operators. Assuming translation invariance, we will drop the superscript $n$. Standard results~\cite{Wolf} show that in the absence of long-range order, $\Psi$ is symmetric if and only if
\begin{align}
\sum_b g_{\theta \, a b} B^b_{\alpha \beta} &= e^{i \nu \theta} \sum_{\alpha', \beta'} U_{\theta\, \alpha \alpha'}^\dagger B^a_{\alpha' \beta'} U_{\theta \, \beta' \beta},\\
\sum_b g_{\mathcal{T}\, a b} \bar{B}^b_{\alpha \beta} &= \sum_{\alpha', \beta'} U^\dagger_{\mathcal{T} \, \alpha \alpha' } B^a_{\alpha' \beta'} U_{\mathcal{T}\, \beta' \beta}. 
\label{eq:transform}
\end{align}
for some unitary $\chi \times \chi$ matrices $U$, and $\nu \in \mathbb{Z}$ is the integer filling of the system.

The transformation laws of Eq.~\eqref{eq:transform} define the $U$ only up to a phase. 
While the onsite symmetries satisfy 
\begin{align}
g_\pi = g_{\mathcal{T}} g^\ast_{\mathcal{T}},
\label{eq:onsite_transform}
\end{align}
since $\mathcal{T}^2 = (-1)^{N_f}$,  the $U$ may realize the symmetry group only projectively,
\begin{align}
U_{\mathcal{T}} U^\ast_{\mathcal{T}} = \omega(\mathcal{T}, \mathcal{T}) U_\pi, \quad \omega \in \textrm{U}(1).
\end{align}
In our case this particular relation is not interesting, as we can choose the phase of $U_\pi$ to set $\omega(\mathcal{T}, \mathcal{T}) = 1$.
However, when we compare the equalities contained in Eqs.~\eqref{eq:transform} and \eqref{eq:onsite_transform}, we obtain 
\begin{align}
g_\pi B = e^{i \nu \pi} U^\dagger_\pi B U_\pi = g_{\mathcal{T}} g^\ast_{\mathcal{T}} B = U^T_{\mathcal{T}} U^\dagger_{\mathcal{T}} B U_{\mathcal{T}} U^\ast_{\mathcal{T}},
\end{align}
which implies
\begin{align}
e^{i \nu \pi} B = B.
\end{align}
Unless $\nu \in 2 \mathbb{Z}$, we arrive at a contradiction.

\section{`Modding-out' and the emergence of Spin and Pin$^{\pm}$ from a lattice model}
\label{sec:aMod}
\subsection{Motivation and setup}
In this appendix, we clarify some subtleties in putting a system on a compact flat manifold $\mathbb{R}^D/\Gamma$. 
To illustrate the problem, let us first explore a more explicit form of the Hamiltonian on $\mathbb{R}^D/\Gamma$ obtained by the modding-out prescription explained in the main text.

To that end, let a subspace $M\subset\mathbb{R}^D$ be a fundamental domain that contains exactly one of the equivalent points of any $\vec{x}\in\mathbb{R}^D$.  Namely, for any $\vec{y}\in\mathbb{R}^D$, there exists a unique $\vec{x}\in M$ and $g\in \Gamma$ such that $\vec{y}=g(\vec{x})$. Note that $g$ is also unique since $\Gamma$ is fixed-point free.  Although it is not necessary, for simplicity we choose $M$ to be simply connected. 

We assume the original Hamiltonian in $\mathbb{R}^D$ is a sum of local terms $\hat{H}=\sum_{\vec{x}}\hat{H}_{A_{\vec{x}}}$, where $\hat{H}_{A_{\vec{x}}}$ has a finite support $A_{\vec{x}}$ around $\vec{x}$.  
Then consider
\begin{equation}
\hat{H}_M=\sum_{\vec{x}\in M}\hat{H}_{A_{\vec{x}}}. 
\label{HM}
\end{equation}
When $\vec{x}$ in this sum is near the boundary of $M$, $\hat{H}_{A_{\vec{x}}}$ might contain operators outside of $M$; in that case they should be understood as the identified operators inside $M$, given by the equivalence relation
\begin{eqnarray}
\label{Oequiv}
\hat{O}_{\vec{x}}^{(i)}\sim_\Gamma \hat{g}\hat{O}_{\vec{x}}^{(i)}\hat{g}^{-1}=\hat{O}_{g(\vec{x})}^{(j)} (U_g)_{ji}.
\end{eqnarray}
Then the natural identification $M\simeq M_\Gamma \equiv \mathbb{R}^D/\Gamma$ allows us to interpret $\hat{H}_M$ as the Hamiltonian on $M_\Gamma$. Note that the infinitely many different choices of $M$ gives the same $M_\Gamma$ as long as one remembers each site in $M_\Gamma$ can be labeled by the labels of any of the equivalent sites in $\mathbb R^D$.

While the choice of $g$ above, as a space group element, is unique, for SOC fermions the symmetry group has to be extended by the $\mathbb Z_2$ fermion parity and $\hat g$ is not uniquely defined. As such, a more careful discussion is warranted and we tackle it in the last subsection of this appendix.

Putting such complications aside, one still sees there are infinitely many valid choices of $M$. The choice of $M$, however, is merely a gauge choice. More precisely, in this appendix we show the following: \emph{given any two valid choices of fundamental domains $M$ and $M'$, the Hamiltonians $\hat H_M$ and $\hat H_{M'}$ are related by a local unitary transformation}.

As an illustrative example, let us consider the following Hamiltonian in 2D:
\begin{equation}
\hat{H}=\sum_{l,m\in\mathbb{Z}}(-1)^l\hat{X}_{(l,m)}\hat{Y}_{(l+1,m)},
\end{equation}
where $X$ and $Y$ are spin operators.
$\hat H$ is invariant under a space group $\mathcal{G}$ generated by glide $G_{x}$,
\begin{eqnarray}\begin{split}
G_x:(l,m)&\mapsto& (l+1,-m),\\
\hat{G}_x\hat{X}_{(l,m)}\hat{G}_x^{-1}&=&\hat{X}_{(l+1,-m)},\\
\hat{G}_x\hat{Y}_{(l,m)}\hat{G}_x^{-1}&=&-\hat{Y}_{(l+1,-m)}.
\end{split}
\end{eqnarray}
and translation $T_y$,
\begin{eqnarray}
\begin{split}
T_y:(l,m)&\mapsto& (l,m+1),\\
\hat{T}_y\hat{X}_{(l,m)}\hat{T}_y^{-1}&=&\hat{X}_{(l,m+1)},\\
\hat{T}_y\hat{Y}_{(l,m)}\hat{T}_y^{-1}&=&\hat{Y}_{(l,m+1)}.
\end{split}
\end{eqnarray}

Let $\Gamma$ be the subgroup of the full symmetry group generated by $G_x^{N_x}$ and $T_y^{N_y}$ with $N_x$, $N_y$ being odd integers. As described, we introduce the equivalence relation $~$:
\begin{eqnarray} 
\hat{X}_{(l,m)}\sim_\Gamma \hat{X}_{(l,m+N_y)}\sim_\Gamma \hat{X}_{(l+N_x,-m)},\\
\hat{Y}_{(l,m)}\sim_\Gamma \hat{Y}_{(l,m+N_y)}\sim_\Gamma -\hat{Y}_{(l+N_x,-m)}.\label{aeqy}
\end{eqnarray}
For example, $M$ can be chosen as $\{(l,m)\,\,|\,\,1\leq l\leq N_x, 1\leq m\leq N_y\}$. Under identification of the edges of $M$ by $\Gamma$, one sees that $M_\Gamma$ is actually the Klein bottle.

As described, the modded-out Hamiltonian is then given by
\begin{eqnarray}
\hat{H}_M&&=\sum_{l=1}^{N_x-1}\sum_{m=1}^{N_y}(-1)^l\hat{X}_{(l,m)}\hat{Y}_{(l+1,m)}\notag\\
&&+\sum_{m=1}^{N_y-1}\hat{X}_{(N_x,m)}\hat{Y}_{(1,N_y-m)}+\hat{X}_{(N_x,N_y)}\hat{Y}_{(1,N_y)},\label{Hxy}
\end{eqnarray}
where we have used Eq.~\eqref{aeqy} to bring $\hat{Y}_{(N_x+1,m)}$ into the region $M$ and $(-1)^{N_x+1}=1$ as $N_x$ is odd.

Staring at Eq.~\eqref{Hxy}, one sees that:
\begin{enumerate}
\item Around the `seam' $l=1$, $\hat{H}_M$ contains couplings (boundary conditions) that do not seem to coincide with the original Hamiltonian defined on the plane.
\item The symmetry of $\hat{H}_M$ is unclear --- does $\hat{H}_M$ still have some reminiscent of the $\mathcal{G}$ invariance of $\hat{H}$?
\end{enumerate}

For the first point,  we argue that the `seam' of $M_\Gamma$ is an artifact of choosing a specific fundamental domain. We can see that the Hamiltonian (deep) inside $M$ [the first sum in Eq.~\eqref{Hxy}] is identical to that of the original Hamiltonian in $\mathbb{R}^2$.  By redefining $M$, any point in $\mathbb{R}^2$ can be taken far away from the boundary of $M$.  For example, for $(l_0,m_0)$ ($l_0,m_0\in\mathbb{Z}$), one may choose
\begin{equation}
M=\{(l,m)\,\,|\,\,|l-l_0| < \textstyle\frac{N_x}{2},|m-m_0| < \textstyle\frac{N_y}{2}\}.
\end{equation}
Therefore, one can argue that any expectation values of local operators or correlation functions around any given points in $M_\Gamma$ are unchanged from those in the bulk (for sufficiently large $N_x, N_y$).  This point can be made clearer using the local unitary equivalence we discuss in the next subsection.

The answer for the second point depends on the property of $\Gamma$. 
To understand this, one should ask: does  $[\vec{x}]=[\vec{y}]$ always imply $[G(\vec{x})]=[G(\vec{y})]$? It is easy to see that this is the case only when $\Gamma$ is a normal subgroup of $\mathcal G$. When $\Gamma$ is normal, the $\mathcal{G}$ invariance of $\hat{H}$ implies the $\mathcal{G}$ invariance of $\hat{H}_M$ (up to local unitaries). Even when $\Gamma$ is \emph{not} normal, however, $\hat H_M$ is still invariant under the normalizer of $\Gamma$ in the full symmetry group (e.g. in the above example, $\Gamma$ is generated by $\hat{G}_x$ only).  For our purpose, the $\mathcal{G}$ invariance of  $\hat{H}_M$ is \emph{not} necessary - the properties that each point of $M_\Gamma$ is locally identical to the bulk and that $\hat{H}_M$ remains TR symmetric are sufficient.

\subsection{Local unitary equivalence}
Here we explicitly show the asserted local unitary equivalence among $\{\hat{H}_M\}$ defined with different fundamental domains, and discuss its physical consequence.

To this end, we first define a slightly more formal, but nonetheless equivalent, procedure to define the modded-out Hamiltonian in Eq.~\eqref{HM}. This merely serves to introduce notations that make the local unitary equivalence manifest.
By definition, for any site $\vec y$ in $\mathbb R^D$ there exists $\vec  x\in M$ and $g_{M}^{\vec  y} \in \Gamma$, both unique, such that $\vec x=g_{M}^{\vec  y} (\vec  y)$. 
In $\mathbb R^D$, the action of $g_{M}^{\vec y}$ on any operator $\mathcal O_{\vec y}$ is represented by first sending $\vec y \rightarrow \vec x$ and then conjugating $\mathcal O_{\vec x}$ by a local unitary operator $\hat s_{M}^{\vec y}$. Note that $\hat s_{M}^{\vec y}$ acts non-trivially only on site $\vec x = g^{\vec y}_M(\vec y)$. After modding out, however, the relabeling $\vec y\rightarrow \vec x$ becomes trivial since $\vec x, \vec  y$ are merely different labels for the same site. The identification of operators, as discussed in Eq.~\eqref{Oequiv}, can then be defined via the map $ \mu_{M}: \mathcal O_{\vec y} \mapsto  \hat s_{M}^{\vec y} \mathcal O_{\vec x} \hat s_{M}^{\vec y \dagger}$.

$\mu_{M}$ can be promoted to a linear map acting on any local operators with support in region $A \subset \mathbb R^D$, provided any two distinct sites in $A $ belong to different equivalence classes. More concretely, we simply have $\mu_{M}( \bigotimes_{\vec  y\in A} \mathcal O_{\vec y}) = \hat{U}_{M}^A \left( \bigotimes_{\vec y\in A} \mathcal O_{g^{\vec y}_M(\vec y)}\right) \hat{U}_{M}^{A \dagger}$ with $\hat{U}_{M}^A = \bigotimes_{\vec y  \in  A } \hat{s}_{M}^{\vec y}$ being a local unitary operator. We can now rewrite the modded-out Hamiltonian in Eq.~\eqref{HM} in a more precise manner:
\begin{align}
\hat{H}_{M} = \sum_{\vec x\in M} \mu_{M} (\hat{H}_{A_{\vec x}} ).
\label{HM2}
\end{align}
In particular, because the elements of $\Gamma$  commute with time-reversal $\mathcal{T}$,  $\hat{ H}_{M} $ is time-reversal symmetric.

Taken at face value, the two Hamiltonians $\hat{H}_{M} $ and $\hat{H}_{M' } $ obtained by modding-out with fundamental domains $M \neq M'$ are not identical. However, the choice of the fundamental domain is only a guage choice and one can always construct a local unitary transform $\hat{\mathcal U}$ such that $\hat{\mathcal U}\hat{ H}_{M} \hat{\mathcal U}^\dagger=\hat{H}_{M'} $. 
In fact, one simply sees that $\hat H_{M'} = \mu_{M'} (\hat H_M)$ and therefore $ \hat{\mathcal U}= \hat U^{M}_{M'}$. (Note again one can freely switch between the equivalent labels for the same site.)

Such local unitary equivalence implies the ground state expectation value of any local operators with support $A_\Gamma \subset M_\Gamma$ is identical to that of the bulk with exponential accuracy. If $A_\Gamma$ corresponds to a region $A$ deep inside $M$, this is evident since the ground state is gapped and the restriction of $\hat H_{M}|_{A_\Gamma}$ is identical to $\hat H|_A$. If $A_\Gamma$ corresponds to a region $A'$ that crosses the border of $M$ (or equivalently the corresponding region is disconnected in $M$), one can nonetheless choose a new fundamental domain $M'$ to center $A'$. In essence, given a `measurement region' $A_\Gamma$, one can move any apparent `seams' arbitrarily far away from it using $\mathcal U$ (bounded by the size of $M_\Gamma$), and hence deduce the measurement outcome is identical to the bulk up to exponential corrections. Important to the discussion is that $\mathcal U$ leaves the region $A_\Gamma$ unchanged. This should be contrasted with a `visible' defect, say an impurity, which can be moved only by invoking translation operator and will therefore shift also the measurement region. 

\subsection{Complications with spinful fermions: Spin and Pin}

As explained in the main text, the `spin' degrees of freedom complicates the situation since the internal rotations or reflections introduce sign factors that depends on the fermion number $\hat{N}$. Here we clarify the mathematical structure of Spin or Pin structure on a compact flat manifold $M_\Gamma=\mathbb{R}^D/\Gamma$ in more detail.

Before discussing (s)pin, let us first examine the Bieberbach group $\Gamma$ itself. The first Bieberbach theorem states that each $g \in \Gamma$ has a unique decomposition into a product of a (possibly fractional) translation $T$ and an orthogonal transformation $B \in O(D)$, $g = T B$.
For our purposes only a discrete subset  $ G \subset O(D)$ will appear, such as reflections and $2 \pi / m $ rotations. To be fancy, we have the short exact sequence
\begin{align}
\xymatrix{1 \ar[r] &  \mathbb{Z}^d \ar[r] & \Gamma \ar[r]^{r}  & G \ar[r] & 1},
\end{align}
where $\mathbb{Z}^d$ denotes the lattice translations in $\Gamma$ and $r$ projects onto the point group part of $g$, i.e., $r(T B) = B$.  The map $r$ is the `holonomy representation' of $M_\Gamma$.  For the Bieberbach manifold $M_\Gamma$, the fundamental group is $\pi^1(M_\Gamma) = \Gamma$; intuitively this is because a straight line connecting two points on $\mathbb{R}^D$ related by $g \in \Gamma$ becomes a closed curve in $M_\Gamma$.
The map $r$ encodes the fact that when traversing a path $g \in  \pi^1(M_\Gamma)$, we return to the point having been `screwed' by the orthogonal transformation $r(g) \in G$.

In the presence of (s)pin degrees of freedom, $G$ is realized as a $\mathbb{Z}_2$ extension encoding the sign structure, which we denote by $G^F$. For instance, for spin-$1/2$, the $2\pi$ rotation is $\hat{C}_{2\pi} = (-1)^{\hat{N}}$, where $\hat{N}$ denotes the fermion number. The screw and glide transformations then inherit these signs, and a $2_1$-screw will square to a translation and the fermion parity, $\hat{S}^2 = (-1)^{\hat{N}} \hat{T}$.
Formally,
\begin{align}
\xymatrix{
1 \ar[r] &  \mathbb{Z}_2 \ar[r] & G^F \ar[r]^{\lambda}  & G \ar[r] & 1},
\end{align}
where $\lambda[(-1)^{\hat{N}}] = 1$.
The most familiar example of this sequence is probably the case of $G=\text{SO(3)}$ and $G^F=\text{Spin}(3)\simeq\text{SU(2)}$.  The choice of $G^F$ depends on the physical model, but for SOC spin-$1/2$ fermions $G^F$ is a discrete subgroup of the `$\text{Pin}^{-}(D)$' group, the unoriented counterpart of $\text{Spin}(D)$.
The `$-$' superscript indicates that mirror reflections satisfy $R^2 =  (-1)^N$, since reflection is implemented on spins as $R_x = i \sigma^x$.
Generally, we can consider both $\text{Pin}^{+}(D)$ and $\text{Pin}^{-}(D)$ depending on whether reflections square to $R^2  = (\pm 1)^N$.

To consistently introduce the equivalence relation on spinful fermions required to mod-out by $\Gamma$, we need a group homomorphism $\epsilon: \Gamma \to G^F$ which encodes how we choose the sign that can accompany each reflection and rotation.
Namely, $\epsilon$ has to satisfy
\begin{equation}
\epsilon(g_1)\epsilon(g_2)=\epsilon(g_1 g_2).
\end{equation}
such that the following diagram commutes:
\begin{align}
\xymatrix{
&   G^F \ar[d]^{\lambda}  \\
\Gamma  \ar[r]^r   \ar[ru]^\epsilon  &G }
\label{eq:pin_structure}
\end{align}
This is a `lift' of the holonomy representation to $G^F$.
Depending on $\Gamma$, it may or may not be possible to find such an $\epsilon$.
In fact, Ref.~\onlinecite{Lutowski2014} shows that Eq.~\eqref{eq:pin_structure} is precisely the condition for the existence of a Spin / Pin structure on $M_\Gamma$.  
All orientable 3-manifolds admit spin structure and all nonorientable 3-manifolds admit $\text{Pin}^-$ structure. Hence, all 10 Bieberbach 3-manifolds (6 orientable and 4 nonorientable) admit Spin or $\text{Pin}^-$ structure, and such $\epsilon$ always exists.

As an illuminating example, let us discuss the space group No.~29. This group is generated by three symmetry elements:
\begin{align}
G_x:\,(x,y,z) &\mapsto (x+\textstyle\frac{1}{2}, -y, z),\\
T_y:\,(x,y,z) &\mapsto (x, y+1, z),\\
S_z:\,(x,y,z) &\mapsto (-x, -y, z + \textstyle\frac{1}{2})
\end{align}
with relations
\begin{align}
G_x^2 = T_x, \quad S_z^2 = T_z, \quad  G_x S_z = T_x S_z G_x,
\label{app:products}
\end{align}
where $T_{x,y,z}$ is translation by one unit. Since $r(G_x)=IC_{2x}$ ($I$ is the inversion) and $r(S_z)=C_{2z}$, we have $\epsilon$ up to signs $\xi_{x,y,z}=\pm1$:
\begin{align}
\epsilon(G_x) &= \xi_x e^{i\frac{\sigma_x}{2 }\pi}= \xi_x i\sigma_x,\\
\epsilon(T_y) &=\xi_y\sigma_0,\\
\epsilon(S_z) &= \xi_z e^{i\frac{\sigma_z}{2 }\pi}= \xi_zi\sigma_z.
\end{align}
For $\epsilon$ to be homomorphic, $\epsilon$ has to keep relations in Eq.~\eqref{app:products}.  
It is easy to see that once we fix $\epsilon(T_x)=\epsilon(T_z)=-1$, all relations are satisfied without specifying $\xi_{x,y,z}=\pm1$.  We thus have $2^3=8$ $\text{Pin}^-$ structures.

Curiously, this group, No.~29, is the single exception that does not admit $\text{Pin}^+$ structure among all Bieberbach groups. For $\text{Pin}^+$ we have to use $\epsilon(G_x)=\xi_x \sigma_x$ so that $\epsilon(G_x^2)=+1$, which also fixes $\epsilon(T_x)=+1$. Then, the third relation in Eq.~\eqref{app:products} is violated.  Therefore, we cannot put particles obeying $\text{Pin}^+$ on the manifold `1st Amphidicosm' obtained by modding-out $\mathbb{R}^3$ by $\Gamma=\text{No.~29}$.  This, however, does not concern us here since we are interested in spin-1/2 electrons.

Note that because we consider models with U(1) symmetry, it seems that we could redefine the reflections by an arbitrary U(1) phase arbitrary phase $\theta$, $R_x \to e^{i \theta N} R_x$.
Choosing then $\theta = \pi/2$ erases the distinction between $\text{Pin}^+$ and $\text{Pin}^-$. 
Formally this amounts to considering U(1), rather than $\mathbb{Z}_2$, extensions of the space group, 
leading to $\text{Spin}_{\mathbb{C}}$ and $\text{Pin}_{\mathbb{C}}$ structures.
However, time-reversal narrows down this U(1) ambiguity since the modding-out process must preserve TR symmetry. 
The space group elements $g \in \Gamma$ and $\mathcal{T}$ commute; the lift $\epsilon$ must  as well, $\epsilon(g) \mathcal{T} = \mathcal{T} \epsilon(g)$, otherwise the `seams' of the region will break TR.
For $S=1/2$ models, this amounts to requiring
\begin{equation}
\sigma_y\epsilon(g)^*\sigma_y=\epsilon(g).
\end{equation}
So TR invariance reduces the U(1) ambiguity back down to $\mathbb{Z}_2$, fixing either $\text{Pin}^{\pm}$.
For example, $\epsilon(G_x)=\xi_xi\sigma_x$ for $\text{Pin}^-$ satisfies this, while $\epsilon(G_x)=\xi_x\sigma_x$ for $\text{Pin}^+$ does not. 
In general, the relation $R^2 (\mathcal{T} R \mathcal{T}^{-1} R^{-1}) = (\pm1)^N$ is invariant under  U(1) redefinitions of $R$ (since $\mathcal{T}$ conjugates phases) and determines which of $\text{Pin}^{\pm}$ to use.

\section{Group-subgroup relations and elementary NS space groups}
As explained in the main text, all 157 NS space groups contains at least one of the elementary NS space groups as $t$-subgroup.
Figure ~\ref{classification} below illustrates which space group contains which elementary space groups, generated based on Ref.~\onlinecite{ITC}.  Elementary NS space groups are grouped into 10 Bieberbach manifolds~\cite{Conway2003}.

Using this group-subgroup relation, one can deduce the constraint on filling $\nu$ for all non-elementary space groups $\mathcal{G}$ based on the result for NS elementary space groups established in the main text. It can in general be expressed as $\nu = m_{\mathcal{G}} n$, where
\begin{eqnarray}
m_{\mathcal{G}}=\max_{\Gamma_0\leq\mathcal{G}}\frac{|\mathcal{G}:T|}{|\mathcal{G}:\Gamma_0|}
\end{eqnarray}
with $\Gamma_0$ being a Bieberbach (fixed-point free) subgroup of $\mG$ and $T$ being the lattice translation subgroup of $\mG$.
(Any Bieberbach group is `elementary' in our notation, but some elementary space groups are not fixed-point free.)
We list the results in Table~\ref{Bbb1}.

\begin{table*}[h!]
\begin{center}
\caption{\label{Bbb1}The filling that may realize sym-SRE phases for nonsymmorphic space groups. Those space groups not listed here are symmorphic and hence {$\nu=2n$}; $n$ in this table is a positive integer $1,2,3,\cdots$.}
\begin{tabular}{|cc|cc|cc|cc|cc|cc|cc|}
No.	&	&No.	&	&No.	&	&No.	&	&No.	&	&No.	&	&No.	&	\\\hline
\sg{4}&$4n$&\sg{39}&$4n$&\sg{66}&$4n$&\sg{100}&$4n$&\sg{129}&$4n$&\sg{165}&$4n$&\sg{201}&$4n$\\
\sg{7}&$4n$&\sg{40}&$4n$&\sg{67}&$4n$&\sg{101}&$4n$&\sg{130}&$8n$&\sg{167}&$4n$&\sg{203}&$4n$\\
\sg{9}&$4n$&\sg{41}&$4n$&\sg{68}&$4n$&\sg{102}&$4n$&\sg{131}&$4n$&\sg{169}&$12n$&\sg{205}&$8n$\\
\sg{11}&$4n$&\sg{43}&$4n$&\sg{70}&$4n$&\sg{103}&$4n$&\sg{132}&$4n$&\sg{170}&$12n$&\sg{206}&$4n^\dagger$\\
\sg{13}&$4n$&\sg{45}&$4n$&\sg{72}&$4n$&\sg{104}&$4n$&\sg{133}&$4n^\dagger$&\sg{171}&$6n$&\sg{208}&$4n$\\
\sg{14}&$4n$&\sg{46}&$4n$&\sg{73}&$4n^\dagger$\tablenote{$\dagger$: $\nu=8n-4$ is prohibited for the noninteracting case~\cite{us3}. There is no known interacting model of sym-SRE at these filling either.}&\sg{105}&$4n$&\sg{134}&$4n$&\sg{172}&$6n$&\sg{210}&$4n$\\
\sg{15}&$4n$&\sg{48}&$4n$&\sg{74}&$4n$&\sg{106}&$4n^\dagger$&\sg{135}&$4n^\dagger$&\sg{173}&$4n$&\sg{212}&$8n$\\
\sg{17}&$4n$&\sg{49}&$4n$&\sg{76}&$8n$&\sg{108}&$4n$&\sg{136}&$4n$&\sg{176}&$4n$&\sg{213}&$8n$\\
\sg{18}&$4n$&\sg{50}&$4n$&\sg{77}&$4n$&\sg{109}&$4n$&\sg{137}&$4n$&\sg{178}&$12n$&\sg{214}&$4n$\\
\sg{19}&$8n$&\sg{51}&$4n$&\sg{78}&$8n$&\sg{110}&$4n^\dagger$&\sg{138}&$8n$&\sg{179}&$12n$&\sg{218}&$4n$\\
\sg{20}&$4n$&\sg{52}&$8n$&\sg{80}&$4n$&\sg{112}&$4n$&\sg{140}&$4n$&\sg{180}&$6n$&\sg{219}&$4n$\\
\sg{24}&$4n$&\sg{53}&$4n$&\sg{84}&$4n$&\sg{113}&$4n$&\sg{141}&$4n$&\sg{181}&$6n$&\sg{220}&$4n$\tablenote{$\nu=4$ is prohibited for the noninteracting case~\cite{us3}. There is no known interacting model of sym-SRE at this filling either.}\\
\sg{26}&$4n$&\sg{54}&$8n$&\sg{85}&$4n$&\sg{114}&$4n$&\sg{142}&$4n^\dagger$&\sg{182}&$4n$&\sg{222}&$4n$\\
\sg{27}&$4n$&\sg{55}&$4n$&\sg{86}&$4n$&\sg{116}&$4n$&\sg{144}&$6n$&\sg{184}&$4n$&\sg{223}&$4n$\\
\sg{28}&$4n$&\sg{56}&$8n$&\sg{88}&$4n$&\sg{117}&$4n$&\sg{145}&$6n$&\sg{185}&$4n$&\sg{224}&$4n$\\
\sg{29}&$8n$&\sg{57}&$8n$&\sg{90}&$4n$&\sg{118}&$4n$&\sg{151}&$6n$&\sg{186}&$4n$&\sg{226}&$4n$\\
\sg{30}&$4n$&\sg{58}&$4n$&\sg{91}&$8n$&\sg{120}&$4n$&\sg{152}&$6n$&\sg{188}&$4n$&\sg{227}&$4n$\\
\sg{31}&$4n$&\sg{59}&$4n$&\sg{92}&$8n$&\sg{122}&$4n$&\sg{153}&$6n$&\sg{190}&$4n$&\sg{228}&$4n^\dagger$\\
\sg{32}&$4n$&\sg{60}&$8n$&\sg{93}&$4n$&\sg{124}&$4n$&\sg{154}&$6n$&\sg{192}&$4n$&\sg{230}&$4n^\dagger$\\
\sg{33}&$8n$&\sg{61}&$8n$&\sg{94}&$4n$&\sg{125}&$4n$&\sg{158}&$4n$&\sg{193}&$4n$& &\\
\sg{34}&$4n$&\sg{62}&$8n$&\sg{95}&$8n$&\sg{126}&$4n$&\sg{159}&$4n$&\sg{194}&$4n$& &\\
\sg{36}&$4n$&\sg{63}&$4n$&\sg{96}&$8n$&\sg{127}&$4n$&\sg{161}&$4n$&\sg{198}&$8n$& &\\
\sg{37}&$4n$&\sg{64}&$4n$&\sg{98}&$4n$&\sg{128}&$4n$&\sg{163}&$4n$&\sg{199}&$4n$& &\\
\end{tabular}
\end{center}
\end{table*}

\begin{figure*}[h!]
\begin{center}
{\includegraphics[width=0.9\textwidth]{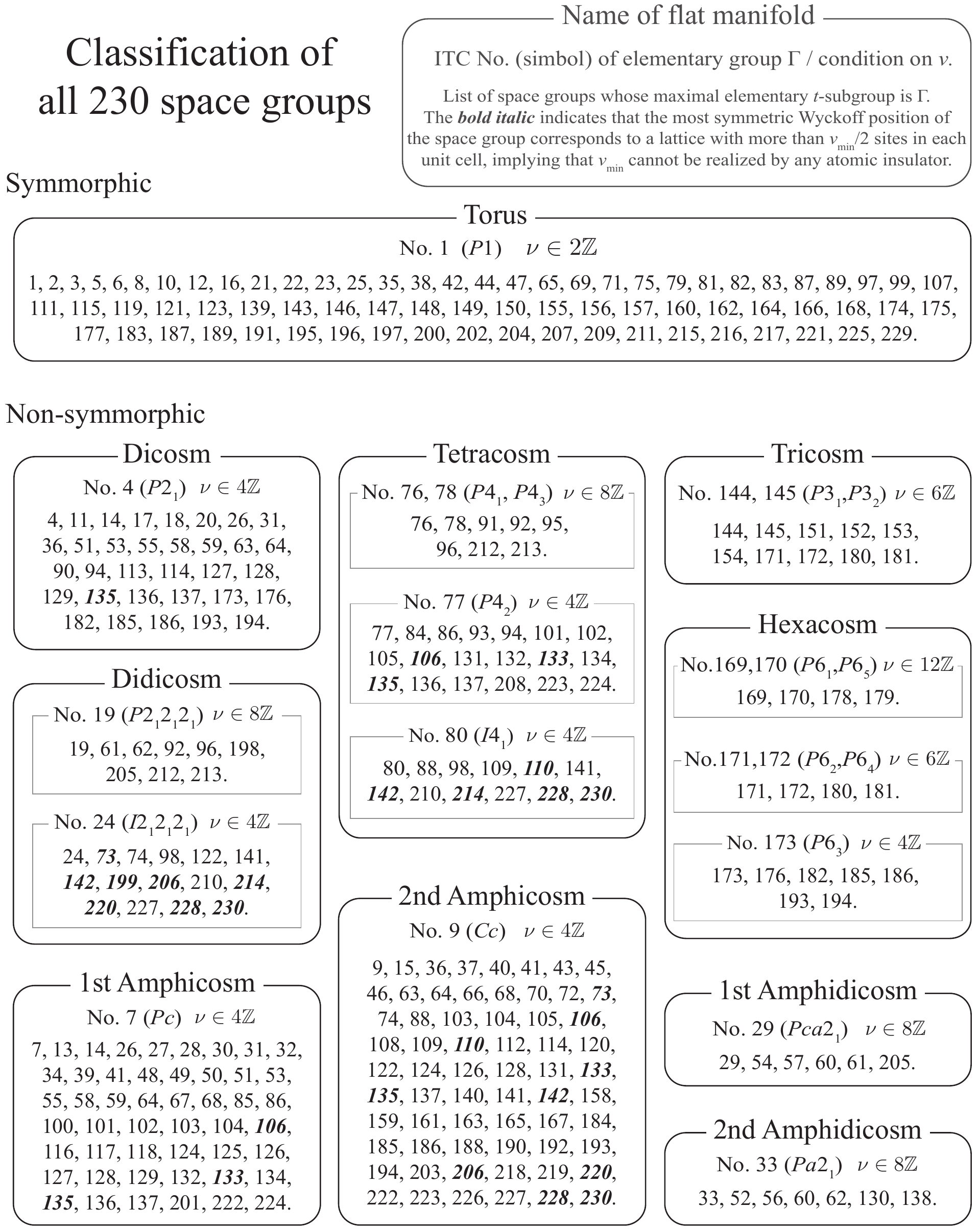}} 
\caption{\label{classification} Classification of all 230 space groups based on its associated Bieberbach manifolds.}
\end{center}
\end{figure*}

\end{document}